\documentclass[epj]{svjour}
\usepackage{graphics}
\begin{document}
\title{Search for scaling onset in exclusive reactions with the lightest nuclei
}
\subtitle{Search for scaling onset}
\author{Yu.N. Uzikov\inst{1}
\thanks{\emph{Present address:} Joint Institute for Nuclear Researches, LNP, Joliott Curie, 6, Dubna, Moscow region, 141980}%
}                     
\offprints{}          
\institute{Joint Institute for Nuclear Researches,  V.P. Dzhelepov Laboratory of Nuclear problems \and
Department of Physics, M.V. Lomonosov Moscow State University}
\date{Received: date / Revised version: date}
%
\abstract{The dimensional scaling  of the differential cross sections of  binary reactions
 $d\sigma/d t\sim s^{-(n-2)}$,  where $n$ is given by  constituent quark counting
 rules, was predicted  for asymptotically high energies  $\sqrt{s}\gg m_i$ and transferred momenta $-t\gg m_i$ at $t/s=const$
 (here $s$ and $t$  are the Mandelstam
 variables and $m_i$  denotes a hadron mass), but
  manifested itself at surprisingly moderate energies
  of few GeV at  large fixed   cms angles $\theta_{cm}\sim 90^\circ$. This behaviour  is observed
  not only in reactions with  free hadrons, but
   with the deuteron  and  $^3He$ too both for electromagnetic and  pure hadronic interactions.
  One may suppose that observed scaling  points out to effective restoration of near-conformal and, probably, chiral symmetry
  in these processes.
   A systematical experimental  study of the scaling   behaviour of the reactions with the deuteron,
  $^3H$, $^3He$,  and $^4He$ nuclei  is   still absent. We consider a possibility to carry out this study 
  in $dd$  collisions  at the JINR Nuclotron.
%
} 
\maketitle
\section{Introduction}
\label{intro}
The structure of the lightest nuclei at short distances ($r_{NN} < 0.5$~fm)
or for high relative momenta ($q > 1/r_{NN}\sim 0.4$~ GeV/c)
constitutes fundamental problem in nuclear physics.
 One of the most important
questions  is the following:
 at which values of the  Mandelstam variables $s$ and $t$
 (or, more precisely, relative  momenta $q$ of nucleons in
  nuclei) does   the transition region
 from the meson-baryon to the quark-gluon picture of nuclei set in?
 So,  the main aim of experiments  on deep inelastic nuclear reactions
 at high transferred momenta in  the so-called {\it cumulative} region
 was to search for dense fluctuations of nuclear matter (fluctons).
 Very interesting features  were observed on this way in inclusive spectra
 which can be interpreted as manifestation of ``drops'' of the quark phase
 in nuclei (see for review  Ref. \cite{Leksin}). However, a quantitative
theory of  cumulative effect is not available
and, therefore, other independent signals for
the transition region are necessary.

 A definite  signature for  transition to the valence  quark region  is given
 by the  constituent
 counting rules (CCR) \cite{mmt,brodskyf}. According to
 the dimensional  scaling \cite{mmt,brodskyf}
 the differential cross
 section  of a binary reaction $AB\to CD$ at  high enough
 incident energy  can be parameterized for a given c.m.s. scattering angle
 $\theta_{cm}$  as
\begin{equation}
\label{general}
\frac{d\sigma}{d\,t}(AB\to CD)= \frac{f(t/s)}{s^{n-2}},
\end{equation}
where
 $n=N_A+N_B+N_C+N_D$ and $N_i$ is the minimum number of
 point-like constituents in the {\it i-th} hadron (for a lepton and photon
one has
 $N_l=1$),
 $f(s/t)$ is a function of  $\theta_{cm}$.
 The CCR follows from a self-similarity hypothesis \cite{mmt} and
perturbative QCD (pQCD) \cite{brodskyf}. The CCR was derived  also from the ADS/CFT duality
\cite{polchinsky}.

  After short review of existing data in sect. \ref{sec2} and discussion in sect.
\ref{sec3} a proposal for new measurements is given in sect. \ref{proposal}.
\begin{figure*}[hbt]
\resizebox{0.75\textwidth}{!}{
 \includegraphics{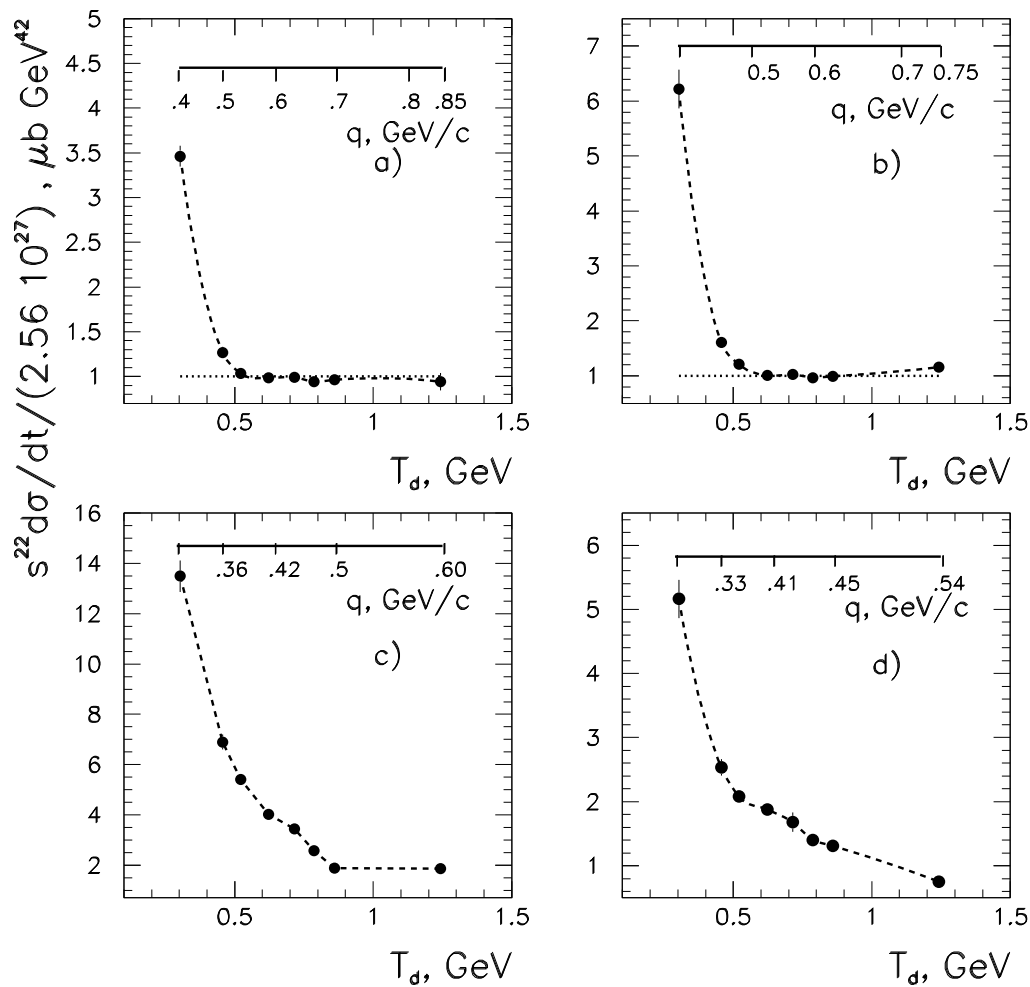}
}
\caption{
The differential
 cross section of the $dd \to~^3Hen$
 and $dd \to~^3Hp$  reactions
 multiplied by $s^{22}$ \cite{uz2006}
 versus the deuteron beam energy at different
 scattering angles:
{\it a} -- $\theta_{c.m.}=60^o $;
 {\it b} -- $50^\circ -52^\circ$;
{\it c} -- $33^\circ -35^\circ$; {\it d} -- $28^\circ$.
 On the upper scale is shown the minimal relative momentum  (GeV/c)
 between nucleons in the deuteron for the ONE mechanism.
The data are taken from Ref. \protect\cite{bizard}.
The stright dotted lines show  the $s^{-22}$ dependence.
 At lower scattering angles
  the plateu is not visible in
 this data, maybe, except for $\theta_{cm}= 33^\circ -35^\circ$. }
\label{fig1}
\end{figure*}

%

\section{Existing data}
\label{sec2}
 Existing  data  at energies about several GeV for
 many  measured  hard  processes
 with free hadrons appear to be consistent with the CCR.
 Among them are reactions of photoproduction of pions,
$\rho$-mesons, kaons on the proton at the photon beam 
 energy $E_\gamma=4 - 7.5$ GeV \cite{andersonccr}, 
$\gamma p$-, $\pi^+ p$-,
$K^+p$- \cite{white}, $pp$-elastic scattering \cite{arndt97} at  large scattering angles $\theta_{cm}\approx 90^\circ$ ,
 strangeness production in the reaction $\pi^-p\to K^0\Lambda$ \cite{kawamura} and others \cite{white}.

 The CCR properties of the reactions with atomic nuclei were observed first in electromagnetic
 interactions with the deuteron.
 So, the deuteron  photodisintegration reaction $\gamma d\to pn$
 follows the $s^{-11}$ scaling behaviour
 at photon energies
  $E_\gamma=1-4$ GeV and large scattering angles $\theta_{cm}\sim
90^\circ$ corresponding to high transversal momenta
 $p_T>1.1 $  GeV/c
\cite{SLAC1,SLAC2,beltz,bochna,Wijesooriya,schulte,mirazita,rossi}.
 Meson-exchange models fail  to explain   the
 $\gamma d\to pn$  data at $E_\gamma >1$ GeV (see, for example,
\cite{bochna}), and therefore several nonperturbative theoretical models
 were suggested \cite{nagorny}-\cite{grishina},\cite{MSU}.
  Since the pQCD is expected to be valid at much higher
 transferred momenta \cite{izgur}, the origin of the observed
 scaling behaviour
 in the reactions with the deuteron at moderate energies is unclear.
 Furthermore, the hadron helicity conservation predicted by the pQCD was not
 confirmed experimentally in the scaling region \cite{Wijesooriya}.
 On the other hand, in these reactions the
 3-momentum transfer $Q > 1$ GeV/c  is  large enough  to probe
 very  short  distances between the nucleons   in nuclei,
 $r_{NN}\sim 1/Q<0.2 \,fm$.
 Presumably,
 nucleons  lose their separate identity
 in this overlapping region and form multi-quark configurations.
 In order to get more insight
 into  the underlying dynamics,
 new measurements were suggested
 to study photodisintegration of the diproton in the $^3He$ at Jlab
\cite{brodsky2004} and the results compatible with the CCR behaviour given by
Eq. (\ref{general})
were obtained   for $^3He(\gamma,pp)n$
 \cite{pomerantz1} and recently
 for $\gamma ^3He\to dp$
\cite{pomerantz2,ilieva}.
\begin{figure}[hbt]
\resizebox{0.43\textwidth}{!}{
 \includegraphics{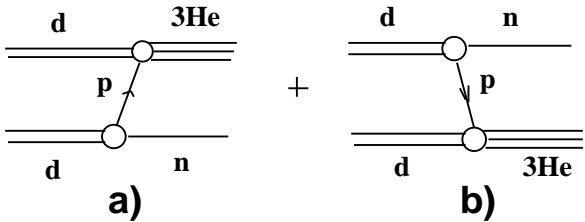}
}
\vspace{0.2cm}
\caption{One nucleon exchange mechanism of the reaction $dd\to ^3He n$.
 This mechanism allows one
 to formulate the scaling onset in terms of internal momenta  $q$ in the deuteron and $~^3He$
 nucleus.}
\label{fig2}
\end{figure}

\begin{figure}[hbt]
\resizebox{0.48\textwidth}{!}{
 \includegraphics{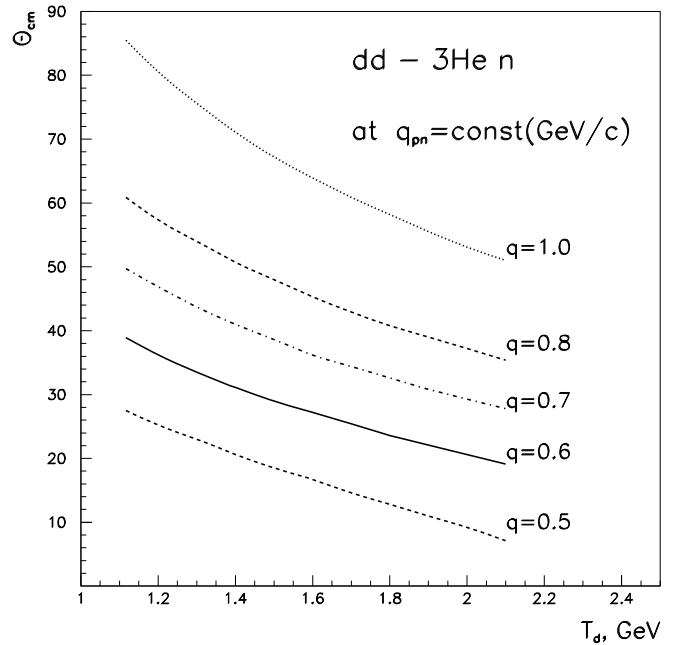}
}
\caption{Internal
 momenta $q_{pn}$
 for the one nucleon exchange mechanism of the reaction
 $dd\to ~^3Hen$ versus the deuteron beam energy $T_d$ for different
 scattering angles $\theta_{cm}$.}
\label{fig3}
 \end{figure}

 Before the CCR behaviour was observed in electromagnetic reactions with the $^3He$ nucleus
 in Refs. \cite{pomerantz1,pomerantz2},
 an indication on the scaling onset in  a pure hadronic reaction, namely the $dd\to~^3Hen$  (or  $dd\to~^3Hp$),
 was  found in  \cite{uz2005} (see Fig. \ref{fig1}).  The key point mentioned in Ref. \cite{uz2005} was
 that in the reaction $\gamma d \to pn$  considered within the  relativistic impulse approximation
 the scaling behavior starts at internal momenta between
 nucleons  in the deuteron
 about $q_{pn}\sim 1$ GeV/c. Considering this value as the true scale of the CCR onset, one can find that
  in the reaction $dd\to~^3Hen$ studied on the basis of the one-nucleon exchange mechanism (ONE)  (see Fig. \ref{fig2})
  this magnitude of the momentum  $q_{pn}$ can be achieved  at beam energies  $T_d\sim 1 $ GeV  for $\theta_{cm}\approx 90^\circ$(see Fig. \ref{fig3}).  There are data on this reaction obtained at SATURNE in 80's
  \cite{bizard}  for  energies  $T_d =0.5 -1.2$ GeV although
   at lower scattering angles $\theta_{cm} \le 60^\circ$.
   Our analysis of these data \cite{uz2005}
   shows (Fig.\ref{fig1}) that in this reaction the CCR behaviour $\sim s^{-22}$ of
   the differential cross section $d\sigma/dt$
   takes the place at $\theta_{cm} = 50^\circ - 60^\circ$
   and  $T_d=0.5-1.25$ GeV  with $\chi_{n.d.f.}^2= 1.18$  (note that in this reaction $n=6+6+9+3=24$).
   In  terms of the internal momentum $q_{pn}$ the scaling
   appears at  $q_{pn} > 0.5$ GeV/c that is twice lower than in the reaction
  $\gamma d\to pn$, where the CCR  behaviour starts at $q_{pn} > 1$ GeV/c. At lower
  scattering angles $\theta_{cm} < 30^\circ$, i.e.
  lower internal momenta $q_{pn}< 0.5$  GeV/c, the CCR behaviour is absent
  as it is expected.
   Up to now, the reaction $dd\to ~^3Hen$ ($~^3Hp$) is the only pure
 hadronic process which involves
  the deuteron and $^3He\, (~^3H)$ nuclei and  found to follow the CCR.

 As shown in \cite{uz2005}, the cross section of
 the reaction $dp\to dp$ also demonstrates the CCR behaviour
 $\sim s^{-16}$   at $T_d= 2T_p = 1-5$ GeV and
 $\theta_{cm}= 120^\circ-130^\circ$.
 However, the
 $\chi^2$-value is not good in this case and, perhaps,
 this is caused by different systematics
 of the data included into analysis \cite{uz2005}.
 For other reactions with the lightest nuclei
 as   $pd\to~^3H \pi^+$, $pd\to~^3He \eta$, $d^3He\to ~^4He p$, a systematic experimental
  study at beam energies
 above 1 GeV and large scattering angles was not done.

\section {Discussion}
\label{sec3}
 From the point of view of constituent quark model, the observed
 $s^{-22}$ behaviour can be considered as a manifestation of the fact
 that all constituent quarks in the initial and final state are
 active in the $dd\to ~^3Hen$ ($~^3Hp$) reaction.
  On the whole, interpretation of such phenomena involves
 the {\it quark-hadron duality}. So, the most sucessfull description of the
 $\gamma d\to pn$ data was achieved within the Quark-Gluon String model
 formulated in terms of Reggeon exchange \cite{grishina}.
 In Ref. \cite{uz2005}
 the Reggeon exchange  model of \cite{grishina}
 with some modifications was also used  to describe
 the $dd\to ~^3Hen$ and $dp\to dp$ reactions in the region of the observed  CCR behaviour.
%

 An exciting feature of existing data on CCR  is the following.
 The scaling behaviour $s^{-11}$ of the cross section of the reaction
 $\gamma d\to pn$ starts at transversal momenta $p_T>1.1$ GeV/c (and $E_{\gamma}> 1 $GeV)
\cite{rossi},
 whereas in the reaction $pp\to d\pi^+$, measured in Ref. \cite{anderson} at the same $p_T =1.0 -1.4$ GeV/c,
 the expected CCR scaling regime $s^{-12}$ is not observed \cite{uz2006}. Why the reaction with photon  differs
 considerably from the reaction with the pion at almost the same kinematic condition?
 As far as we know, the answer to this question is absent. An attempt to explain this 
 difference will be done below.
 Furthermore, in the reaction
 $dd\to ~^3Hp$ the $s^{-22}$ behaviour is appeared \cite{uz2005}
 at lower transversal momenta  $p_T>0.6 $ GeV/c. A  similarly low boundary in
 terms  of the beam energy
 and internal momenta is found  for the scaling onset $\sim s^{-17}$ in the reaction
 $\gamma ~^3He\to pd$ \cite{pomerantz2}.
 On the other hand, the scaling behaviour $s^{-11}$ of the diproton photodisintegration in the reaction
 $^3He(\gamma,pp)n$
 starts  at photon energy $E_\gamma\sim 2$ GeV  \cite{pomerantz1}, that
 is twice larger than the scaling onset in the reaction  $\gamma d\to pn$.
Thus, search for scaling onset in  reactions with different nuclei is the first 
task in this study.

 Empirical success of the  CCR  at moderate energies and transferred momenta
 is hardly related to non-perturbative QCD and
 is considered in current literature as a manifestation of near-conformal
 symmetry of ``the  physical QCD'' \cite{brodsky-teramod}. Indeed, classical Lagrangian of
 QCD has scale (conformal)
  and chiral invariance in the limit of zero quark
 masses. Both these symmetries are broken by quantum effects. As a result,
  the effective coupling constant $\alpha_s$ depends on  transferred momentum squared $Q^2$,  and
 due to spontaneously broken chiral symmetry constituent quarks become  dynamical
 masses. There are some arguments, however,  that at rather low  transferred
 momenta $Q^2<1$
 (GeV/c)$^2$ the  running coupling constant $\alpha_s(Q^2)$ is large 
and approximately independent of
 $Q^2$ \cite{deur}.
 In this region, in order to get an
 effective restoration of conformal invariance and explain the CCR onset,
 one needs a mechanism,
 which would provide a reduction of constituent quark masses for light flavors.
    Following the  mechanism of effective restoration of chiral
 symmetry for excited hadrons  proposed by Glozman \cite{Glozman}, we assume here that
 for central collisions
 in nuclear reactions  rather high excitations (as compared to $\Lambda_{QCD}\sim 200-300$ MeV)
 of hadrons/nuclei are possible in intermediate states. Therefore, intermediate
 hadronic system follows  quasi-classical approximation when
 quantum effects are irrelevant to internal dynamics of hadrons. In this
 regime, the dynamical quark masses are reduced \cite{Glozman} that could lead (at $\alpha_s(Q^2)
 \approx const$) to effective restoration of conformal symmetry, and, as a result, to
 the CCR behaviour of  enough hard nuclear reactions. In addition,  partial restoration of
 the chiral symmetry means decreasing the $\pi NN$  coupling constant \cite{Glozman}, that  allows
 one to explain absence of the CCR behaviour in the $pp\to d\pi^+$ reactions \cite{uz2006}.
 If chiral symmetry is completely restored in a given excited nucleon it cannot decay 
 into  the $\pi N$ channel. 
 On the other hand, this mechanism   suggests that
 in the $pp\to d\rho^+$ reaction the CCR occurs, because the $\rho$- meson is not a Goldstone boson \cite{Glozman}.
 It is important  to check this  difference between the $pp\to d\pi^+$ and  
$pp\to d\rho^+$
 in experiment.  One should note that a similar result would appead due to 
 real restoration of conformal symmetry too. Indeed, the nuclear matter density in
 the short-range
 configurations with high internal momenta between nucleons $q\sim 1$ GeV/c
 probed in these reactions,
 is close to the critical one, $\varepsilon_c\sim 1$GeV/fm$^3$,
 that corresponds to the phase transition \cite{emeliyanov}.

  Therefore, an important task
 is to search for scaling  behaviour
 of different  exclusive reactions with the lightest nuclei
 at large transversal momenta $p_T>0.6$ GeV/c.
 The nearest task is to determine the boundaries of the scaling region
 of the $dd\to~^3He n$ ($dd\to~^3H p$) reaction which was studied in \cite{bizard}
 by completely different  motivation and very poore statistics was obtained
 in that experiment as compared to the $\gamma d\to pn$ reaction in the scaling region
\cite{SLAC1,SLAC2,beltz,bochna,schulte,mirazita,rossi}.
 Within the ONE mechanism (Fig.\ref{fig2}), as seen from Fig.\ref{fig1},
 the CCR behaviour in the reaction
 $dd\to ~^3He\,n$ starts at rather small  internal momenta in the deuteron $q\approx0.5$ GeV/c.
 It is important to verify this scale at other kinematical
 conditions, i.e. at higher beam energies and lower angles or at lower energies
 and higher angles.
 New data  are necessary
  to   check  whether the CCR behaviour
  is valid in the reaction $dd\to ~^3He\,n$  at (i)  energies $T_d>1.2$ GeV
  for $\theta_{cm}>60^\circ$
 and (ii)  scattering angles $\theta_{cm}=60^\circ -90^\circ$ at $T_d>0.5
 $GeV, corresponding
 to   relative momenta in the deuteron $q>0.8$ GeV/c (Fig. \ref{fig3}).
 Basically, one has to know what is the real parameter for the
 scaling regime -- either internal nucleon momentum $q$\footnote{ Of course,
 this parameter can be  determined only  within a certain mechanism of the reaction.
 The  ONE mechanism is used here as the most natural one.}, or the transversal
 momentum $p_T$.

\section{Proposal for JINR Nuclotron}
\label{proposal}
 We propose to measure the cross section
 of reaction $dd\to$ $~^3Hp$ (or $dd\to~^3He\,n$)  at  energies 1 -2.5 GeV and
 $\theta_{cm}\leq 50^\circ$  using the
  BM@N    facility at the JINR Nuclotron.
 The proposed measurement will allow one (i) to check
 whether the CCR behaviour
 of the reaction $dd\to ~^3H\,p$  is valid at these kinematical conditions
 which correspond to high internal momenta $q \geq 0.6$ GeV/c, but
 were not accessible at SATURNE \cite{bizard}
 and (ii) determine the boundaries of this
 behaviour in terms $\theta_{cm},\, p_{T}$ or $q_{pn}$.
 This measurement was proposed \cite{loi2008} first  for WASA-at-COSY \cite{wasa},
 however, not performed. We present below the estimations of
 the counting rate obtained in Ref. \cite{loi2008}.

\subsection{Counting rate esimation}

\label{counting-rate}
The  counting rate at beam energies 1.2$ - $2.2 GeV is estimated using
  the  SATURNE  data \cite{bizard}.
 Assuming the scaling law $s^{-22}$,
the differential cross section at fixed angle $\theta_{cm}$
was extrapolated  to  higher energies
in the laboratory system as
\begin{eqnarray}
\label{cm2lab}
 \frac{d\sigma}{d\sigma_{lab}}=
 J\,\frac{\pi}{p_ip_f}\,\frac{d\sigma}{dt}=
 \frac{J}{J_{0}}\,\frac{p_ip_f}{p_i^0p_f^0} \frac{s_0^{22}}{s^{22}}
\Biggl (\frac{d\sigma}{dt}\Biggr)_0,
\end{eqnarray}
 where $p_i$ $(p_f)$, and $s$  are the initial (final)
 cms momentum and the invariant mass squared, respectively,  at given $T_d$, whereas
 $p_i^0$ $(p_f^0)$ and $s_0$ are the corresponding  values at $T_0=1.243$ GeV that is the maximal
 beam energy in the experiment \cite{bizard}.
 The  Jacobian of transformation $J$ from the cms to lab-system
  is given in the Table \ref{tab:rate1}.
 The counting rate for the angular interval
$ \theta_{lab}=16^{\circ}-17^\circ$ ( i.e. $\theta_{cm} =48^\circ-50^\circ$)
was calculated under assumption that the cms  differential cross
section at
$T_d=1.234$ GeV and $\theta_{cm}=50^\circ$ is equal to
$\Bigl (\frac{d\sigma}{dt}\Bigr)_0=(0.3\pm0.02 )\mu b/$GeV$^2$ \cite{bizard}.
\begin{table}[h]
\begin{center}
\caption{Counting rate $N$ (for 24 hours) expected at $\theta_{lab}=16^\circ-17^\circ$
for the solid angle $2\pi\,
\sin{(17^\circ)}\pi/180=0.03$rad,
 estimated using   data  \cite{bizard}  for the
 luminosity $L=1.35\times 10^{30} cm^{-2}sec^{-1}$.
 }
\vspace{1mm}
\label{tab:rate1}
\begin{tabular}{cccccc}     \hline\noalign{\smallskip}
$T_d$, GeV &  $\kappa=\frac{\pi}{p_i\ p_f}$, & $J$ &
$\frac{p_ip_f}{p_i^0p_f^0} \frac{s_0^{22}}{s^{22}}$     &
 $d\sigma/d\Omega_{lab},$ & $N$ \\
& GeV$^{-2}$ & &  & nb/sr
&   \\
 \hline
1.117 &    3.32 &   7.92    & 1   &   &    4320 \\
1.243 &  2.9  &  8.0& 0.64 & 827    & 2770\\
1.3 &  2.84&  8.03 & 0.52  & 675     & 2255\\
1.4 &  2.64&  8.09 & 0.36 & 470    & 1570\\
1.5 &  2.46 &  8.15& 0.26 & 342    & 1140\\
1.6 &   2.3 &  8.21 &0.18 & 239     & 795\\
1.7 &  2.16 &  8.27& 0.13 & 174    & 575\\
1.8 &  2.0  &  8.33 & 0.091& 122   & 405 \\
1.9 &  1.93 &  8.4& 0.065&  88     & 280 \\
2.0 & 1.83 &   8.46  & 0.046 & 63    & 205\\
2.1 &  1.73 &  8.52 &  0.033& 45   & 150\\
2.2 &  1.65 &   8.58 & 0.024& 33  &  108\\
\noalign{\smallskip}\hline
\end{tabular}
\end{center}
\end{table}

 Therefore, for the solid
 angle $\Delta\Omega_{lab}=2\pi\,
 \sin{\theta_{lab}}\Delta \theta_{lab}=0.03$ rad  the number of events expected at $T_d=2.2$ GeV
  during 24 hours
  will be equal to $N=108$ at the luminosity   $L=1.35\times 10^{30}
 cm^{-2}sec^{-1}$ (see Table \ref{tab:rate1}).
 Note, the   acceptance and corrections for registration efficiency
  were not taken into account in this estimation.
 In total, in order to measure the $d\sigma/d\Omega_{cm}$ at
 $\theta_{cm}=50^\circ$, for six energies 1.3, 1.5, 1.6, 1.8, 2.0 and 2.2 GeV  with statistical
 uncertainty
 $\leq 4$\% ($N\geq 600$ events)
 we need  two weeks.
  With increasing beam energy and $\theta_{cm}$ the required beam time is
 quickly increased.

\section{Conclusion}
\label{conclusion}
 Search for transition region  from hadron
 to quark/gluon degrees of freedom in nuclei
 is an  important task in the QCD treatment of nuclear structure and reactions.
 Constituent  Counting Rules, derived from
 the perturbative QCD, when
 being applied for exclusive reactions with lightest nuclei at high $Q^2$,
 can give a definite signal for transition to the valence quark region.
 Some of the binary reactions with the deuteron or/and $^3He$ follow the CCR
 form of Eq. (\ref{general}) at energies 1 - 2 GeV at large scattering angles.

 However, the perturbative QCD, as a basis of CCR,
 can be hardly valid at these  rather low energies and momenta.
  New data on these and other
 exclusive reactions are required to clarify the underlying CCR dynamics.
 One of the most important task is determination of boundary of scaling
 domain for different reactions.
 JINR Nuclotron facility provides an unique
 possibility to study these properties in $dd$ and $pd$ collisions
 above 1 GeV.

  Measurement of  energy dependence of the differential cross
 section of the reaction $dd\to~^3He\,n$  in the interval $T_d=1 - 2.2$ GeV
 at fixed cms scattering angles
 within    $\theta_{cm}=30^\circ-50^\circ$, corresponding to the expected
 boundary between scaling and non-scaling behaviour for this reaction, is
  considered here.
 This proposal can be considered  as a first step of a broad experimental
 program, which will include other exclusive  reactions with lightest nuclei:
 $dd \to dd$, $pd\to~^3H\pi^+$, $pd\to ~^3He\eta$,
 $dd\to ~^4He\eta$, $pp\to d\rho^+$, $p~^4He\to p~^4He$, $p~^4He\to d~^3He$.

 I would like to  thank Yu.~Petukhov and V.~Kurbatov for help 
in estimation of the counting rate and N.N.~Nikolaev for usefull discussions.


\end{document}